\documentclass[conference]{IEEEtran}
\IEEEoverridecommandlockouts
\usepackage{cite}
\usepackage{amsmath,amssymb,amsfonts}
\usepackage{algorithmic}
\usepackage{graphicx}
\usepackage{multirow}
\usepackage{booktabs}
\usepackage[acronym]{glossaries}
\usepackage{textcomp}
\usepackage{subfig}
\usepackage{xcolor}
\usepackage{amsmath}
\usepackage{hyperref}
\usepackage{bm}
\def\BibTeX{{\rm B\kern-.05em{\sc i\kern-.025em b}\kern-.08em
    T\kern-.1667em\lower.7ex\hbox{E}\kern-.125emX}}
\begin{document}

\title{Improved Disentangled Speech Representations Using Contrastive Learning in Factorized Hierarchical Variational Autoencoder
\thanks{The work of Yuying Xie is supported by China Scholarship Council.}
}

\author{\IEEEauthorblockN{Yuying Xie, Thomas Arildsen, Zheng-Hua Tan}
\IEEEauthorblockA{Department of Electronic Systems, Aalborg University, 
Denmark \\
{yuxi@es.aau.dk, tari@its.aau.dk, zt@es.aau.dk}}}

\maketitle

\newacronym{fhvae}{FHVAE}{factorized hierarchical variational autoencoder}
\newacronym{vae}{VAE}{variational autoencoder}

\begin{abstract}
Leveraging the fact that speaker identity and content vary on different time scales, \acrlong{fhvae} (\acrshort{fhvae}) uses different latent variables to symbolize these two attributes.
Disentanglement of these attributes is carried out by different prior settings of the corresponding latent variables.
For the prior of speaker identity variable, \acrshort{fhvae} assumes it is a Gaussian distribution with an utterance-scale varying mean and a fixed variance. 
By setting a small fixed variance, the training process promotes identity variables within one utterance gathering close to the mean of their prior. 
However, this constraint is relatively weak, as the mean of the prior changes between utterances.
Therefore, we introduce contrastive learning into the \acrshort{fhvae} framework,  to make the speaker identity variables gathering when representing the same speaker, while distancing themselves as far as possible from those of other speakers. 
The model structure has not been changed in this work but only the training process, thus no additional cost is needed during testing.
Voice conversion has been chosen as the application in this paper.
Latent variable evaluations include speaker verification and identification for the speaker identity variable, and speech recognition for the content variable.
Furthermore, assessments of voice conversion performance are on the grounds of fake speech detection experiments.
Results show that the proposed method improves both speaker identity and content feature extraction compared to \acrshort{fhvae}, and has better performance than baseline on conversion. 
\end{abstract}


\section{Introduction}
\label{sec: intro}
Disentangled representation learning~\cite{bengio2013representation, polyak21_interspeech, aloufi2020privacy} tries to extract features for representing the independent data attributes separately. 
Even though unsupervised learning is a hot topic nowadays, paper~\cite{locatello2019challenging} revealed that unsupervised disentangled representation learning is fundamentally impossible, and thus supervised or weakly supervised data are generally needed.
Applications of disentangled representation learning are broad, and one in speech is voice conversion~\cite{FHVAE, zhao2022disentangling,chan2022speechsplit2, qian2019autovc, tang2022avqvc, Reinhold2021contrastive, lian2022robust}. 

The basic strategy of using disentangled representation learning on voice conversion contains two stages: training and conversion.
Training makes neural networks learn to extract features to represent speaker identity and content separately.
To make the generated audio sound like target speaker but retain same content from source speaker utterance, the converted utterances are generated from the speaker embedding of the target speaker and the content embedding from the source speaker. 
Speaker embedding and content embedding should contain as much corresponding information as possible but no information overlap, so as to get desired conversion performance.
This strategy has been applied extensively in recent work~\cite{qian2019autovc, tang2022avqvc, lian2022robust, Reinhold2021contrastive}. However, most work has requirements on data.
Specifically, AutoVC~\cite{qian2019autovc} carried out embedding disentanglement under an auto-encoder framework, by controlling the bottleneck dimension carefully to allow only content information to pass through. Speaker embedding in this work is extracted from a pre-trained speaker encoder according to~\cite{wan2018generalized}, and the segment length for extraction is between 3.5s to 4.5s. 
The authors of~\cite{tang2022avqvc} use an encoder and a codebook for content embedding extraction, while speaker embedding is got as the difference between the encoder output and the codebook output. Trimming the silence of speech with a fixed threshold is an essential pre-processing in this work.
The authors of~\cite{Reinhold2021contrastive} used a deterministic style encoder and a statistical content encoder to extract speaker embedding and content embedding separately. Contrastive predictive coding is used on style encoder output and the framework is trained end-to-end. The segment length is required between 2s and 4s.

\Acrfull{fhvae}~\cite{FHVAE} assumes  that speaker identity changes between sequences (longer than 200ms), while content varies between segments (200ms in experiments). 
Compared with other works, \acrshort{fhvae} neither requires lengthy utterances or pre-processing, nor a pre-trained model for extracting speaker embeddings.
However, \acrshort{fhvae} has its own limitations.
To equip sequential latent variables with representation capability,
the prior adopts a Gaussian distribution with mean which changes between utterances and a fixed small variance which makes sequential latent variables are similar within the utterance. However, the relationship of sequential variables between utterances has not been considered.

Contrastive learning intends to make representations symbolising the same class become as similar as possible, while making representations symbolising different classes become as dissimilar as possible~\cite{hadsell2006dimensionality, he2020momentum, sarkar2019time}. This idea fits speaker embedding learning, and there also exists some work using contrastive learning in speaker embedding extraction, like~\cite{tang2022avqvc, wan2018generalized}.

Inspired by the idea behind contrastive learning and related work, this paper applies contrastive learning in the \acrshort{fhvae} framework. 
Compared with \acrshort{fhvae} only concerning information within one utterance, contrastive learning brings cross-utterance information during training to enhance latent variable representation capability. 
The proposed method does not change the framework (and thus does not increase the model complexity during test), but the training strategy. Speech recognition and speaker verification are used to analysis the extracted latent variables and fake speech detection for converted utterances.

\section{Factorized Hierarchical Variational Autoencoder}
\label{sec:original}

The nature of speech shows that speaker identity and content vary on different time scales: speaker identity changes between sequences, while content varies faster, within segments. 
To utilize this fact in speech signal disentangled representation, \acrshort{fhvae} assumes two variables follow a sequence-dependent prior and a sequence-independent prior respectively. Latent variables representing speaker identity and linguistic content in the graphical model are denoted sequential variable and segmental variable in the following.

Specifically, we denote the speech feature (e.g. log-magnitude spectrogram) dataset $\bm{D}=\{{\bm{X}}^ {(i)}\}_{i=1}^M$, where $i$ represents the $i$-th sequence in the dataset, and the dataset contains $M$ utterances.
Assume each utterance ${\bm{X}}^{(i)}$ contains $N^{(i)}$ segments: ${\bm{X}}^{(i)}=\{{\bm{x}}^{(i,n)}\}_{n=1}^{N^{(i)}}$. 
$z_1^{(i,n)}$ and $z_2^{(i,n)}$ denote segmental latent variable and sequential latent variable in order, and superscript $(i,n)$ expresses that the variables are for the $n$-th segment of the $i$-th utterance. The following notation will omit the superscript $(i)$ for simplicity as training happens on segment-scale. 
Besides, $p(\cdot)$ and $q(\cdot)$ represent prior and posterior distributions, while $\theta$ and $\phi$ represent parameters in the generative model and the inference model, respectively.

The generative model in \acrshort{fhvae} assumes data $\bm{X}$ for each sequence is generated in the following process: (1) latent variable $\bm{\mu}_2$ is drawn from prior $p_{\theta}(\bm{\mu}_2) = \mathcal{N}(\bm{\mu}_2|\bm{0},\bm{I})$; (2) sequential latent variables $\{\bm{z}_2^{(n)}\}_{n=1}^N$ and segmental latent variables $\{\bm{z}_1^{(n)}\}_{n=1}^N$ are drawn from priors $p_{\theta}(\bm{z}_2|\bm{\mu}_2) = \mathcal{N}(\bm{z}_2|\bm{\mu}_2, \sigma_{\bm{z}_2}^2\bm{I})$ and $p_{\theta}(\bm{z}_1)=\mathcal{N}(\bm{z}_1|\bm{0}, \bm{I})$. 

The inference model in \acrshort{fhvae} assumes all posteriors of latent variables: $q_{\phi}(\bm{\mu}_2^{(i)})$, $q_{\phi}(\bm{z}_2|\bm{x})$ and $q_{\phi}(\bm{z}_1|\bm{x}, \bm{z}_2)$ are Gaussian distributions.
Means and variances of $q_{\phi}(\bm{z}_2|\bm{x})$ and $q_{\phi}(\bm{z}_1|\bm{x},\bm{z}_2)$ are from the neural network, while the mean $\bm{\widetilde \mu}_2^{(i)}$ of $q_{\phi}(\bm{\mu}_2^{(i)})=\mathcal{N}(\bm{\mu}_2^{(i)}|\bm{\widetilde \mu}_2^{(i)},\bm{I})$ is regarded as a parameter in the inference model.

The structure of \acrshort{fhvae} contains three modules: encoder 1 (denoted $\rm{Enc_1}$ in the following) and encoder 2 (denoted $\rm{Enc_2}$) for extraction of latent variables $\bm{z}_1$ and $\bm{z}_2$, and decoder (denoted $\rm{Dec}$) for data reconstruction and conversion. The data flow of the \acrshort{fhvae} framework is shown as below:
\begin{align}
\bm{z}_2 &= {\rm{Enc_2}} ({\bm{x}})
\label{z2 in FHVAE}\\
\bm{z}_1 &= {\rm{Enc_1}} (\bm{x}, \bm{z}_2)
\label{z1 in FHVAE}\\
\bm{ y} &= {\rm{Dec}} (\bm{z}_1, \bm{z}_2)
\label{x in FHVAE}
\end{align}
and $\bm{y}$ denotes the generated data.

The objective function of \acrshort{fhvae} contains four terms: log-likelihood loss to measure the reconstruction performance; the KL divergence to calculate the distance between prior and posterior of $\bm{z}_1$ and $\bm{z}_2$; the log-likelihood loss of mean of $\bm{z}_2$. The mathematical formulation is:
\begin{equation}
\begin{aligned}
\label{eq:FHVAE}
    &\mathcal{L}_{orig}^{(i,n)} \\&=  \mathbb{E}_{q_{\phi}(\bm{z}_{1}^{(i,n)},\bm{z}_{2}^{(i,n)}|\bm{x}^{(i,n)})}\left[\log p_{\theta}(\bm{y}^{(i,n)}|\bm{z}_1^{(i,n)},\bm{z}_2^{(i,n)})\right] \\
    & - \mathbb{E}_{q_{\phi}(\bm{z}_2^{(i,n)}|\bm{x}^{(i,n)})}\left[D_{KL}(q_{\phi}(\bm{z}_1^{(i,n)}|\bm{x}^{(i,n)},\bm{z}_2^{(i,n)})\|p_{\theta}(\bm{z}_1^{(i,n)}))\right]  \\
    & - D_{KL}(q_{\phi}(\bm{z}_2^{(i,n)}|\bm{x}^{(i,n)})\|p_{\theta}(\bm{z}_2^{(i,n)}| \bm{\Tilde{\mu}}_2^{(i)})) \\
    &  + \frac{1}{N^{(i)}}\log p _{\theta}(\bm{\Tilde{\mu}}_2^{(i)})
\end{aligned}
\end{equation}

When applied in voice conversion, superscript $src$ and $tar$ denote the latent variables from source speaker and target speaker in the following, and superscript $con$ denotes the variables prepared for voice conversion.
For sequential latent variable $\bm z_2$, the new latent variable $\bm{z}_2^{con}$ is generated by shifting the mean of $\bm{z}_2$ from $\bm{\mu}_2^{src}$ to $\bm{\mu}_2^{tar}$:
\begin{equation}
    \bm{z}_2^{con} = \bm{z}_2^{src} - \bm{\mu}_2^{src} + \bm{\mu}_2^{tar} 
\end{equation}
And the converted utterance is generated as:
\begin{equation}
    \bm{y}^{con} = {\rm {Dec}}(\bm{z}_1^{src}, \bm{z}_2^{con})
\end{equation}
The decoder uses the the new sequential latent variable $z_2^{con}$ and the segmental latent variable $z_1^{src}$ from the source speaker to generate the converted utterance.

\section{Proposed method}
\label{sec:proposed method}



As shown in Section~\ref{sec:original}, \acrshort{fhvae} assumes sequence-dependent and sequence-independent prior for sequential latent variable $\bm{z}_2$ and segmental latent variable $\bm{z}_1$, respectively. 
As linguistic content changes between segments but its statistic is global for all sequences, segmental-scale variable $\bm{z}_1$ has sequence-independent prior. 
For sequential latent variable $\bm{z}_2$, the mean $\bm{\mu}_2$ of its prior $p_{\theta}(\bm{z}_2|\bm{\mu}_2)$ is assumed drawn from a standard Gaussian distribution for each utterance, i.e. $\mu_2$ changes between utterances. 
Thus prior $p_{\theta}(\bm{z}_2|\bm{\mu}_2)$ is sequence-dependent.
The training target for the sequential latent variable is to make $\bm{z}_2$ become close to $\bm{\mu}_2$, and to other $\bm{z}_2$ from the same utterance in Euclidean space. 
This is carried out by setting a small fixed variance in the prior
$p_{\theta}(\bm{z}_2|\bm{\mu}_2) = \mathcal{N}(\bm{z}_2|\bm{\mu}_2, \sigma_{\bm{z}_2}^2\bm{I})$, where
 $\sigma_{\bm{z}_2}$ equals 0.5 as in~\cite{FHVAE}.

However, 
only  encouraging latent variable $\bm{z}_2$  close within an utterance is relatively weak. The relationship of $\bm{z}_2$ between sequences should also be considered.
Therefore, we introduce contrastive learning 
to improve sequential representation. 
The idea behind contrastive learning is to make representations more similar within the same class, and less similar between  classes. And thus the cross-utterance information is tapped.  

\begin{figure}[t]
  \centering
  \includegraphics[width=\columnwidth]{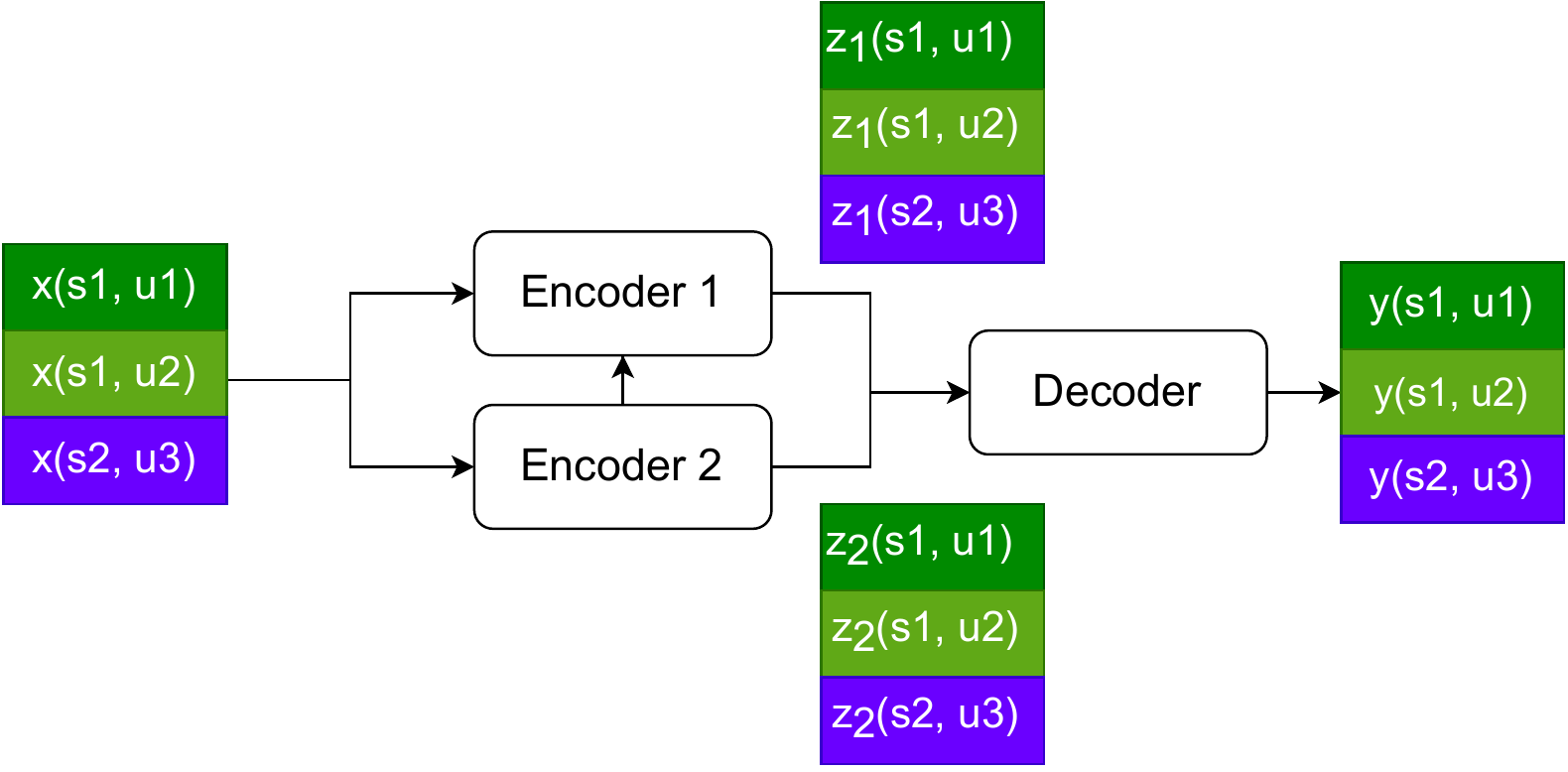}
  \caption{Framework of the proposed method. Three different utterances are fed in the framework during every training step for contrastive learning.}
  \label{fig:contrastiveFHVAE}
\end{figure}

The framework of the proposed method is shown in Fig.~\ref{fig:contrastiveFHVAE}. Same with \acrshort{fhvae}, this framework contains three modules: encoder 1 for segmental variable extraction, encoder 2 for sequential variable extraction, and decoder for reconstruction and conversion. The difference between the proposed method and \acrshort{fhvae} is: for every training step, the input contains speech features of three different utterances from two speakers: utterances 1 and 2 are from speaker 1, and their speech features denote $\bm{x}(\mathrm s1, \mathrm u1)$ and $\bm{x}(\mathrm s1, \mathrm u2)$ in Fig.~\ref{fig:contrastiveFHVAE}; utterance 3 is from speaker 2, and is denoted $\bm{x}(\mathrm s2, \mathrm u3)$. $(\mathrm si, \mathrm uj)$ denotes the feature from the $j$-th utterance of speaker $i$ in the following.


\begin{figure}[t]
  \centering
  \includegraphics[width=0.45\columnwidth]{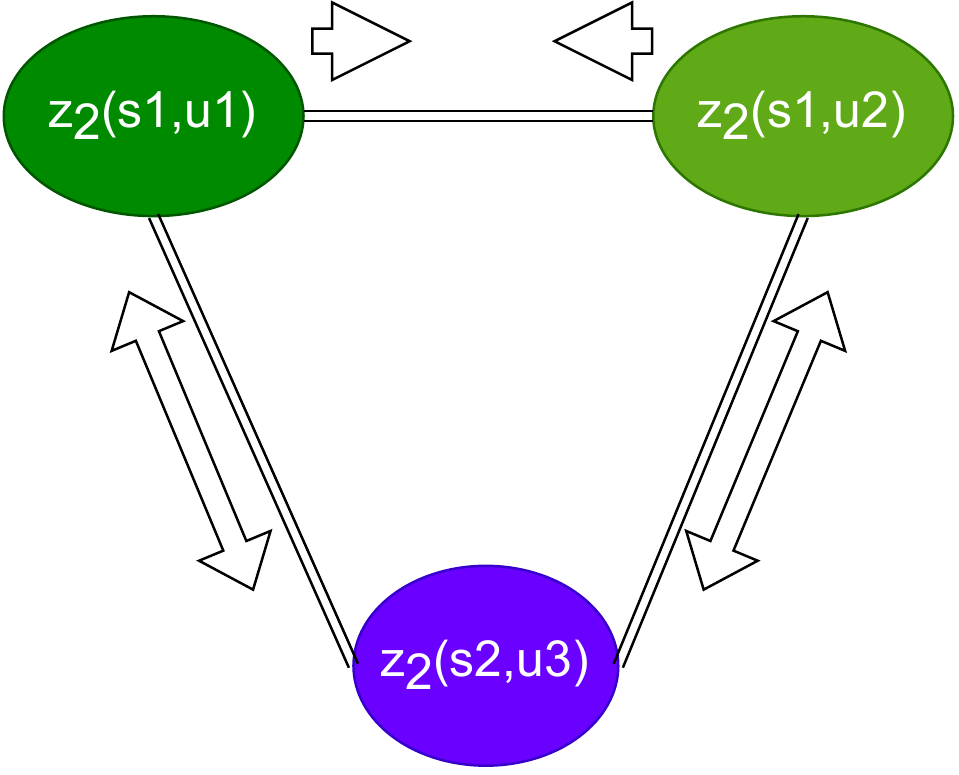}
  \caption{Conceptual illustration of using contrastive learning on latent variable $\bm{z}_2$.}
  \label{fig:contrastive illustration on z2}
\end{figure}

According to the data properties, the sequential latent variable $\bm{z}_2$ for utterances 1 and 2 should be naturally closer since they represent the same speaker. The sequential variable $\bm{z}_2$ for utterance 3 should be further from the other two since it represents another speaker. This enables contrastive learning for improving performance of latent variable $\bm{z}_2$ on speaker identity representation. The graphical illustration of using contrastive learning on  $\bm{z}_2$ is shown in Fig.~\ref{fig:contrastive illustration on z2}. The training target  is to decrease the distance between $\bm{z}_2(\mathrm s1, \mathrm u1)$ and $\bm{z}_2(\mathrm s1, \mathrm u2)$, and increase the distances between $\bm{z}_2(\mathrm s2, \mathrm u3)$ and the two former. In this work, $L^2$-norm is used as distance metric. The contrastive loss for $\bm{z}_2$ is shown as below:
\begin{equation}
\label{eq:proposed method contrastive}
\begin{aligned}
    \mathcal{L}_{cont} =\ & 
    \lambda \| \bm{z}_2(\mathrm s1,\mathrm u1) - \bm{z}_2(\mathrm s1,\mathrm u2) \|_2^2 \\
    & - \beta\| \bm{z}_2(\mathrm s1,\mathrm u1) - \bm{z}_2(\mathrm s2,\mathrm u3) \|_2^2  \\
    & - \beta \| \bm{z}_2(\mathrm s1,\mathrm u2) - \bm{z}_2(\mathrm s2,\mathrm u3) \|_2^2
\end{aligned}
\end{equation}
Based on preliminary experiments and to make positive pair and negative pairs have same contribution in training, $\lambda = 0.01$, and $\beta = 0.005$ in this work.

The total loss function of the proposed method is:
\begin{equation}
\label{eq:proposed method}
    \mathcal{L}_{total} =  
    \mathcal{L}_{orig} - \mathcal{L}_{cont}
\end{equation}
in which $\mathcal{L}_{orig}$ is the loss for \acrshort{fhvae} as shown in~\eqref{eq:FHVAE}.

The voice conversion process is the same as for \acrshort{fhvae} which is explained in Section~\ref{sec:original}.

\section{Experiments}
\label{sec:experiments}
\subsection{Dataset}
\label{subsec:dataset}
\noindent\textbf{TIMIT:} The training set, development set and core test set of TIMIT have been used, each containing respectively 462, 50, and 24 speakers. 
8 utterances (labeled with 'SI' and 'SX') per speaker are used in the experiments. Log-magnitude spectrogram is used as input feature, with the window size and hop size equal to 25ms and 10ms.
For conversion, one utterance is chosen randomly under label 'SI' for each speaker in test set, namely 24 content-different utterances are chosen. Conversion has been made pair-to-pair, and thus 576 converted utterances have been generated.

\noindent\textbf{VCTK:} VCTK dataset contains 110 speakers (62 females and 48 males). We randomly selected utterances from 88, 11 and 11 speakers as training set, development set and test set without overlap. As to conversion, 2 utterances are randomly selected for each speaker in test set, and 462 converted utterances are generated.  

\begin{table}
  \caption{Latent variable evaluation.}
  \label{tab:latent variable}
  \centering
  \begin{tabular}{ c| c| c| c | c}
    \hline
    \multirow{3}{*}{\textbf{Framework}} & \multicolumn{3}{c|}{\textbf{Sequential}} & \textbf{Segmental} \\
    \cline{2-5}
    & \multirow{2}{*}{\small{EER(\%)}} &  \multicolumn{2}{c|}{\small{Accuracy(\%)}} & \multirow{2}{*}{\small{WER(\%)}}\\
    \cline{3-4}
    & & \scriptsize{GRU} & \scriptsize{GRU+FC} \\ 
    \hline
    FHVAE & $3.13$ & $79.69$ & $93.23$ & $27.5$ ~~~ \\ \hline
    proposed & $\bm{2.73}$ & $\bm{85.94}$ & $\bm{96.88}$ & $\bm{26.3}$ ~~~ \\ \hline
  \end{tabular}
\end{table}

\begin{table*}
  \caption{Voice conversion evaluation}
  \label{tab:voice conversion}
  \centering
  \begin{tabular}{ c| c| c| c | c| c}
    \hline
    \multirow{2}{*}{\textbf{Framework}} & \multicolumn{5}{c}{\textbf{fake speech detection}} \\
    \cline{2-6}
    & \scriptsize{F2F} & \scriptsize{M2M} & \scriptsize{F2M} & \scriptsize{M2F} & \scriptsize{All} \\
    \hline
    FHVAE (TIMIT) & $0.767$ & $0.719$ & $0.721$ & $0.701$ & $0.713$  ~~~ \\ \hline
    proposed (TIMIT) & $\bm{0.772}$ & $\bm{0.721}$ & $\bm{0.727}$ & $\bm{0.705}$ & $\bm{0.717}$ ~~~ \\
    \hline
    \hline
     AUTOVC (VCTK) & $0.638$ & $0.672$ & $0.666$ & $0.664$ & $0.648$ ~~~ \\ \hline
    proposed (VCTK) & $\bm{0.680}$ & $\bm{0.713}$ & $\bm{0.679}$ & $\bm{0.685}$ & $\bm{0.678}$ ~~~ \\ \hline
  \end{tabular}
\end{table*}

\subsection{Baseline}
The proposed method is compared with the original FHVAE framework, on latent variable extraction and voice conversion capability. Besides, AutoVC~\cite{qian2019autovc}, as a state-of-the-art work mentioned in~\ref{sec: intro}, has been chosen as another baseline for comparison on conversion quality. The released model and vocoder from the authors of AutoVC have been used for converted utterances generation.

\subsection{Implementation details}
\label{subsec:implementation details}
\noindent \textbf{Framework structure and training details: }The structure of encoders and decoders are the same as in \acrshort{fhvae} and in the proposed method. All encoders and decoders contain one LSTM layer with 256 units and a fully-connected layer. The dimensions of latent variables in both frameworks equal 32. For fair comparison, the training settings are also same for baseline and the proposed method. Batch size equals 768. Each batch contains data from three kinds of utterance equally in the proposed method and random data for baseline. Learning rate equals $10^{-4}$, and the optimizer is Adam~\cite{ADAM}. Vocoder used in this work is HifiGAN~\cite{kong2020hifi}.

\begin{figure*}[htbp]
    \centering
    \subfloat[FHVAE]{
    \label{fig1:a}
    \includegraphics[width=0.28\paperwidth]{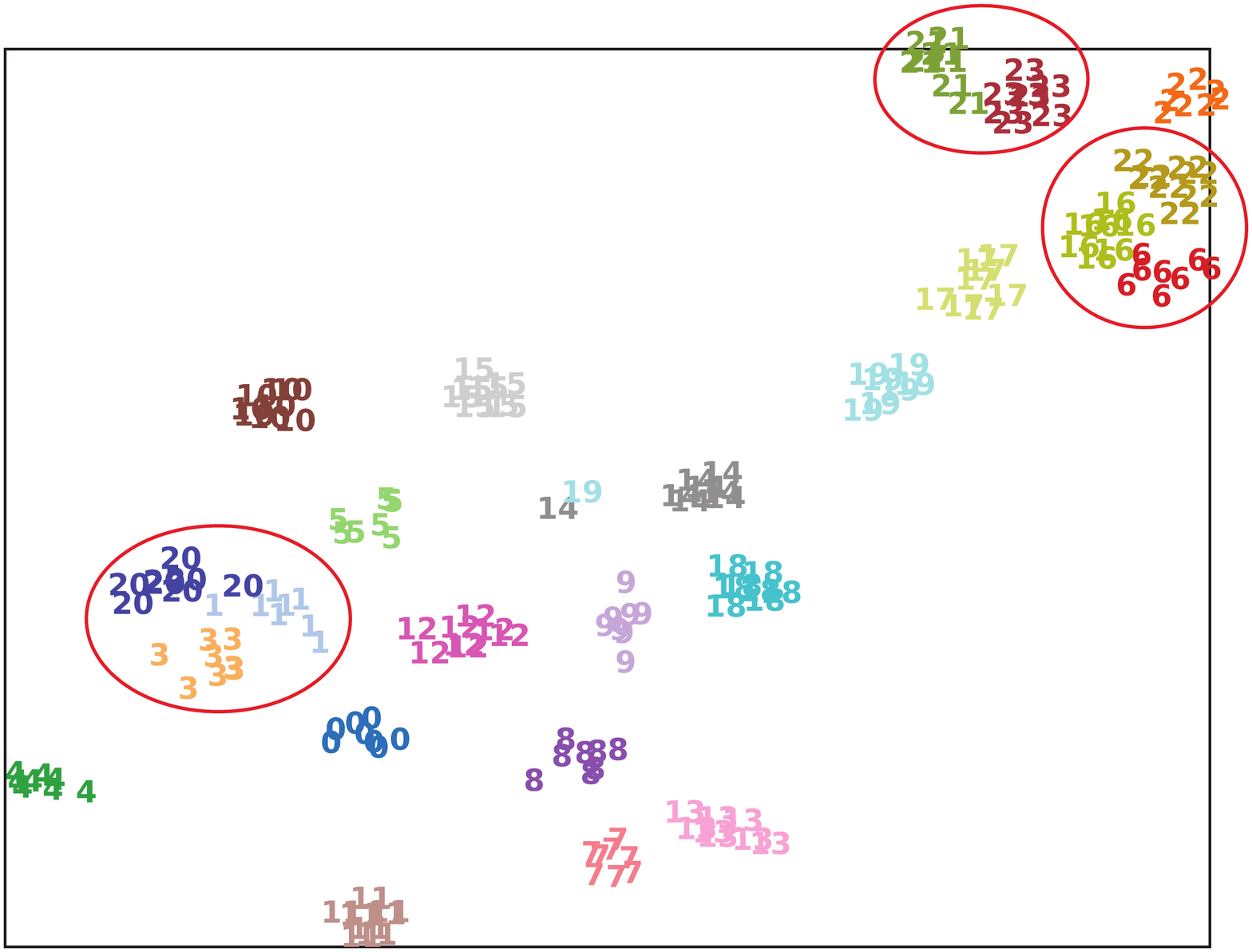}
    }
    \subfloat[The proposed method]{
    \label{fig1:b}
    \includegraphics[width=0.28\paperwidth]{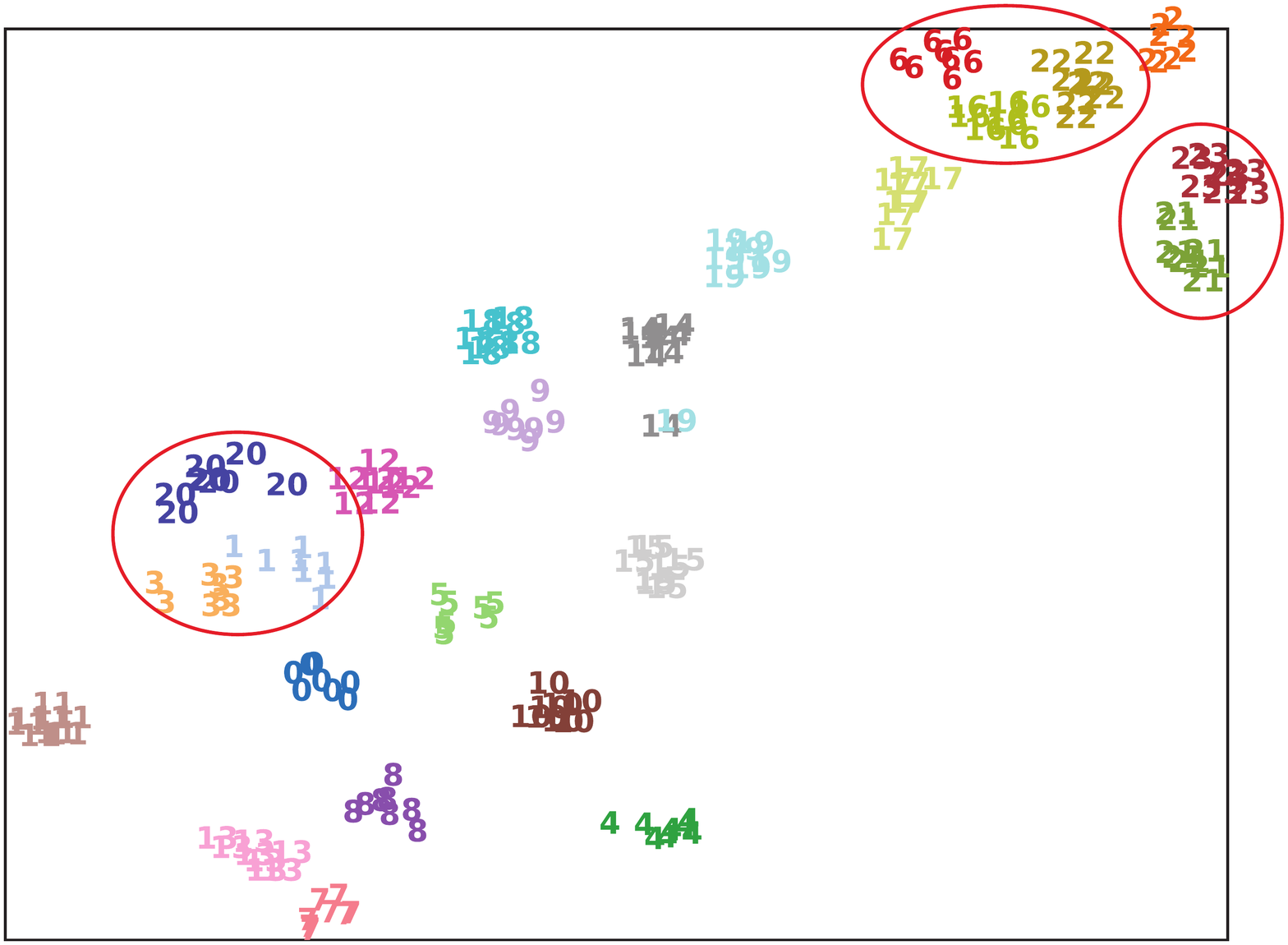}
    }
    \caption{t-SNE plot of extracted sequential features from the compared frameworks. All parameter settings for 
both t-SNE plots are the same. The red circles indicate that the cluster dispersion is more obvious in the proposed method.}
    \label{t-SNE}
\end{figure*}

\noindent \textbf{Segmental-level feature evaluation}: The mean and variance of latent variable $z_1$ are used as the extracted segmental-level feature in the evaluation experiment. 
As the segmental-level feature assumed to represent linguistic content, speech recognition is used for evaluation. The speech recognition system is implemented by Kaldi toolbox~\cite{povey2011kaldi} and experiment details are the same as in~\cite{xie2021disentangled}. The results are shown in Table~\ref{tab:latent variable} under header 'Segmental'. 'WER' denotes word error rate (WER), and lower WER indicates better performance.

\noindent \textbf{Sequential-level feature evaluation}: 
The mean of $\bm{z}_2$ has been used as the sequential-level representation.
Speaker verification and speaker identification experiments have been done for fairness.
For speaker verification, the equal error rate (EER) based on cosine similarity is displayed in Table~\ref{tab:latent variable} under the header 'EER(\%)', the lower the better. Two neural network architectures have been applied for speaker identification: one implemented by 1-layer GRU and the other by 1-layer GRU with 512 units followed by a dense layer.
The details of speaker identification are the same as in~\cite{xie2021disentangled}. The results from the two classifiers can be found in Table~\ref{tab:latent variable} with header 'Accuracy(\%)', under 'GRU' and 'GRU+FC' respectively. The higher accuracy means the better performance.

\noindent \textbf{Voice conversion evaluation}: 
Conversion evaluation is based on fake speech detection, using the open toolkit \href{https://github.com/resemble-ai/Resemblyzer}{\textit{Resemblyzer}}\footnote{https://github.com/resemble-ai/Resemblyzer} for speech quality and similarity evaluation. Results are shown in Table~\ref{tab:voice conversion} under 'fake speech detection'. Higher score shows better performance. Specifically, fake speech detection has been done between intra-sexual conversion which shown with header 'F2F', 'M2M', and inter-sexual conversion under 'F2M' and 'M2F'. The overall evaluation results are shown under header 'All'.
\href{https://yuxi6842.github.io/contrastiveFHVAE.github.io/#traditional}{Demos}\footnote{https://yuxi6842.github.io/contrastiveFHVAE.github.io/\#traditional} could be found on our website.


\subsection{Results and analysis}
\label{subsec:results and analysis}
\noindent \textbf{Sequential-level feature:}
Table~\ref{tab:latent variable} shows consistent improvement from the proposed method in all experiments. Additionally, t-SNE~\cite{van2008visualizing} plots give visualization in Fig.~\ref{t-SNE}. Each number $k$ in Fig.~\ref{t-SNE} denotes one utterance from the $k$-th speaker. The clusters become more separated between different speaker classes in the proposed method. For instance, features from speaker 6, 22 and 16 look a little bit more separated in Fig.~\subref*{fig1:b}, so do cluster 21 and 23.
Cluster 20 and 1, 3 show more obviously the advantage of introducing contrastive learning to the FHVAE framework.

\noindent  \textbf{Segmental-level feature:} The results  in Table~\ref{tab:latent variable} show that the introduced contrastive learning not only improves sequential latent variable extraction, but also slightly improves segment latent variable extraction, indicating better disentanglement.

\noindent  \textbf{Voice conversion:} Results shown in Table~\ref{tab:voice conversion} indicate that speech converted by the proposed method exhibits slight improvement on fake speech detection compared with the original FHVAE. It proves that contrastive learning strategy helps slightly on voice conversion performance of FHVAE. Besides, the experiment results also show that the proposed method works better than AutoVC.

\section{Conclusion}
\label{sec:conclusion}
As the constraint of sequential latent variable is relatively weak in \acrshort{fhvae}, we introduce contrastive learning to improve it. 
The proposed method not only considers the distance of sequential variables within one utterance, but also among utterances through contrastive learning. 
No more layers but only the learning strategy has been changed. 
Experiment results show that, compared with baseline, the proposed method improves both sequential latent variable and segmental latent variable extraction performance; meanwhile in voice conversion application, the proposed method also shows better performance.

\bibliographystyle{IEEEbib}
\bibliography{ref}

\end{document}